\documentclass[manuscript=article]{achemso}

\usepackage[version=3]{mhchem} 
\usepackage[T1]{fontenc}       

\usepackage{subfig}

\author{Alexander Galstyan}
\affiliation[UCL]{Institute of Condensed Matter and Nanosciences, Universit\'e Catholique de Louvain, 2 chemin du cyclotron, BOX L7.01.07, B1348 Louvain-la-Neuve, Belgium}
\email{alexander.galstyan@uclouvain.be}
\author{Yuri V. Popov}
\affiliation{Joint Institute for Nuclear Research, Dubna, Russia}
\alsoaffiliation{Skobeltsyn Institute of Nuclear Physics, Lomonosov Moscow State University, Moscow, Russia}
\author{No\"el Janssens}
\affiliation[UCL]{Institute of Condensed Matter and Nanosciences, Universit\'e Catholique de Louvain, 2 chemin du cyclotron, BOX L7.01.07, B1348 Louvain-la-Neuve, Belgium}
\author{Francisca Mota-Furtado}
\affiliation[RHUL]{Department of Mathematics, Royal Holloway, University of London, Egham, Surrey TW20 0EX, United Kingdom}
\author{Patrick F. O'Mahony}
\affiliation[RHUL]{Department of Mathematics, Royal Holloway, University of London, Egham, Surrey TW20 0EX, United Kingdom}
\author{Piero Decleva}
\affiliation[UTR]{Dipartimento di Scienze Chimiche e Farmaceutiche, Universita' di Trieste, Trieste, Italy}
\author{Nicola Quadri}
\affiliation[UTR]{Dipartimento di Scienze Chimiche e Farmaceutiche, Universita' di Trieste, Trieste, Italy}
\author{Bernard Piraux}
\affiliation[UCL]{Institute of Condensed Matter and Nanosciences, Universit\'e Catholique de Louvain, 2 chemin du cyclotron, BOX L7.01.07, B1348 Louvain-la-Neuve, Belgium}

\title[H$_2$O ionisation in SAE]
  {Ionisation of H$_2$O by a strong ultrashort XUV pulse: a model within the single active electron approximation}

\abbreviations{SAE,TDSE,DFT}

\begin{document}







\begin{abstract}
We present and discuss a new computationally inexpensive method to study, within the single active electron approximation, the interaction of a complex system with an intense ultrashort laser pulse. As a first application, we consider the one photon single ionisation of the highest occupied molecular orbital of the water molecule by a laser pulse. The ionisation yield is calculated for different orientations of the molecule with respect to the field polarization axis and for different carrier envelope phases of the pulse, and compared against predictions of another single active electron approach.
\end{abstract}

\section{Introduction}

The recent development  of coherent sources of light such as attosecond lasers \cite{Krausz2009}, high-order-harmonic generation (HOHG) sources \cite{Sansone2006,Goulielmakis2008} or free-electron lasers \cite{Ackermann2007,Elettra} has opened the route to the study of the interaction of matter with intense femtosecond and even sub femtosecond radiation pulses in the XUV regime. Such studies allow one to analyze electron dynamics in atoms or electron and nuclei dynamics in molecules with an unprecedented degree of temporal and spatial resolution. Within this context, the interaction of the water molecule with such radiation pulses is of particular relevance in medical physics, in particular, radiotherapy, since water is one of the main components of most  living tissues.\\

In the present contribution, we develop a new computationally inexpensive method to study, within the single active electron (SAE) model and in the non-relativistic regime, the single ionisation of atoms and molecules by an intense femtosecond or sub femtosecond XUV pulse. Specifically we aim to address the total ionisation yield as function of field frequency, intensity and orientation dependence of the polarisation vector with respect to the molecule. We apply this approach to the water molecule in its ground state while paying attention to the validity of the assumptions we make in such a treatment. We assume that the pulse duration is small compared to the characteristic time of the vibrational motion of the molecule and that, during the interaction, the geometry of the molecule is not modified as we employ the fixed nuclei approximation. 

Recently, by measuring the ratio of H$_2$O and D$_2$O high harmonic yields, Farrell {\it et al.} \cite{Farrell2011} managed to characterize the nuclear motion in the mole\-cular  states of H$_2$O$^+$. They showed that by contrast to the ionisation of the highest occupied molecular orbital (HOMO), the single ionisation of the second least bound orbital HOMO-1 triggers a fast nuclear dynamics of the molecular ion through a strong bending motion of the molecule. As a result, we only consider here, the single ionisation of the orbital HOMO 1{\it b}$_1$ which leaves the geometry of the molecule practically unchanged during the interaction with the pulse. In fact, the period of the fastest oscillation in the water molecule, namely the asymmetrical stretching of the OH bonds, is 8.9 fs \cite{Huber1979} which is much longer than the pulse durations we consider here. In other words, we can assume that the molecule is frozen during its interaction with the pulse. However, it is important to note that experimentally, it is impossible to know, {\it a priori},  from which orbital the electron is ejected. Farrell's results show that ionisation from HOMO-1 3a$_1$, which sends the molecule into the $\tilde{A}$ $^2A_1$ state of H$_2O^+$, strongly excites the bending mode at photon energies around 14.8 eV. This puts some limitation on the photon energy used in the present work and requires us to pay attention to the bandwidth of the pulse.\\

Our approach is based on a model that was first developed to treat the interaction of atomic hydrogen with an electromagnetic pulse \cite{Tetchou2011}. We work in the momentum space and use the velocity gauge. The main idea is to replace the kernel of the Coulomb potential by a sum of $N$ symmetric separable potentials, each of them supporting a bound state of the system. This method, which we denote by SPAM for Separable Potentials for Atoms and Molecules, allows one to reduce the 4-dimensional time-dependent Schr\"odinger equation (TDSE) to a system of  $N$ 1-dimensional Volterra integral equations depending only on time. As a result, the integration over the spatial coordinates which, in some cases, requires prohibitively large grids or bases, is completely avoided. Each separable potential may be calculated from the exact wave function of the atomic state it supports. However, its analytical expression is not always unique. We developed a procedure to calculate these separable potentials. It provides results for the electron energy spectra that compare very well with  those obtained by solving the TDSE with the exact Coulomb potential in situations where the number of essential atomic states playing a significant role is low. By moving from the momentum space to the configuration space, it is easy to show that the separable potentials have a finite range. Let us note that once the separable potentials are determined, the continuum states are automatically defined and, being solutions of the same equation as the one satisfied by the exact bound states taken into account in the calculations, they are orthogonal to these bound states. To generalize to more complex systems such as the water molecule, we proceed along the same lines. We first generate the HOMO in terms of gaussian type orbitals by means of the well established quantum chemistry software package GAMESS(US). It is then straightforward to move to the momentum space and to define the corresponding separable potential which is unique in this case.  As for atomic hydrogen, the final step involves solving a 1-dimensional Volterra integral equation. For the sake of completeness, it is important to mention that our approach is not gauge invariant as it is the case for most of the approximate treatments. The problem of the gauge and the delicate question of the possible existence of a privileged gauge are discussed in detail in the context of the present model by Galstyan {\it et al.} \cite{Galstyan2017}.\\

The problem of the interaction of the water molecule with a femtosecond or sub femtosecond  XUV pulse has almost never been treated theoretically up to now. As far as we know, most of the theoretical calculations have been performed at a wave length of 800 nm. However, it is interesting from the methodological point of view to mention three different contributions. In the first one,  Borb\'ely {\it et al.} \cite{Borbely2010} study the ionisation of the water molecule by an intense ultrashort half-cycle electric pulse. They performed both quantum mechanical and classical calculations  within the single active electron and frozen core approximation. They considered the ejection from the HOMO 1b$_1$. Since this orbital is mainly constructed from the $2p_z$ orbital of the oxygen atom when the molecule is lying in {\it x-y} plane, they modelled it by an hydrogenic $2p_z$ orbital with an effective charge chosen to reproduce the ionization energy of the HOMO. They found good agreement between the classical and quantum mechanical calculations at high field intensity where the over-the-barrier ionisation mechanism is dominant. In the second contribution, Della Picca {\it et al.} \cite{DellaPica2012} study the orientation-dependent single ionisation of fixed-in-space H$_2$O by a short laser pulse for two wave lengths: 800 nm and 76 nm well in the XUV regime. In their calculation, which is based on the strong field approximation (SFA)\cite{Galstyan2016} , the initial and final states are described by single-Slater determinants of spin-orbitals, the spatial part of it being calculated by means of the same quantum chemistry software package as in our case. They take into account the five occupied molecular orbitals. The SFA is a first order theory in the sense that the ionisation results from the absorption of a "virtual" photon that is supposed to describe tunneling emission. In the high frequency regime, we have shown \cite{Galstyan2016} that the SFA gives qualitatively good results. In their contribution, Della Picca {\it et al.} showed that the HOMO-1 dominates the single-electron emission process when the laser is polarized along the symmetry axis of the molecule and that the electron emission is in general favored in the direction along the laser polarization direction. In the third contribution Petretti {\it et al.} \cite{Petretti2013} apply the single active electron approximation time dependent Schr\"odinger equation (SAE-TDSE) method to the water molecule. They solve a 4-dimensional TDSE within the single-active-electron approximation to treat the orientation-dependent ionisation of H$_2$O in few-cycle 800 nm linear-polarized laser pulses. The molecular orbitals are Kohn-Sham orbitals obtained by using the LB94 exchange-correlation functional. They showed that the HOMO dominates the overall ionsation behaviour except in the nodal plane of this orbital where the dominant contribution comes from the HOMO-1. The role of the carrier envelope phase is also investigated. \\

Our contribution is organized as follows. After this  introduction, we present our method. First, very briefly in the case of atomic hydrogen and then, in more detail, in the case of the water molecule. In the third section, we first present some tests of our method in the case of atomic hydrogen. Before the conclusions and perspectives we present our results for the water molecule and compare them against the predictions of SAE-TDSE method described in detail by Petretti {\it et al}\cite{Petretti2013}. 

Atomic units combined with the Gaussian system for the electromagnetic field are used throughout unless otherwise specified.\\
\section{Theoretical model}

In this section, we define the pulse and describe our model. For the sake of clarity, we first consider briefly the case of atomic hydrogen. Details of the calculations are found in Galstyan {\it et al.} \cite{Galstyan2017}. We then show, in more detail, how it can be generalized to a more complex system such as the water molecule.

\subsection{Description of the pulse}

We use the dipole approximation and assume that the electric field is linearly polarized along the unit vector $\mathbf{e}$, the direction of which, unless explicitly stated, coincides with our z-axis. The case of an arbitrary polarization does not introduce any additional complication and does not require more computer resources. The electric field is defined in terms of the vector potential $\mathbf{A}(t)$ as follows:
\begin{equation}
\mathbf{A}(t)=A(t)\mathbf{e}=A_0\sin^2\left[\frac{\pi t}{T}\right]\sin(\omega t+\phi)\mathbf{e},\;\;\;\;\;\;\; 0\le t\le T,
\end{equation}
where $\omega$ is the radiation frequency, $\phi$, the carrier envelope phase and $T$, the full pulse duration. In term of the peak intensity, the amplitude $A_0$ is given by:
\begin{equation}
A_0=\frac{1}{\omega}\sqrt{\frac{I}{I_0}},
\end{equation} 
where $I_0=3.5\times 10^{16}$ W/cm$^2$ is the atomic unit of intensity. Before considering the atomic hydrogen case, it is convenient to define the following quantity:
\begin{equation}
b(t)=-\frac{1}{c}\int^t_0A(\xi)\mathrm{d}\xi,
\end{equation}
where $c$ is the speed of light.

\subsection{Atomic hydrogen}
We work in the momentum space  and use the velocity gauge. Under these conditions, the TDSE that describes the interaction of atomic hydrogen with the radiation pulse is:
\begin{eqnarray*}
\left [\mathrm{i}\frac{\partial}{\partial t}-\frac{p^2}{2}-\frac{1}{c} A(t)(\mathbf{e}\cdot\mathbf{p})-
\frac{1}{2c^2}A^2(t)\right]\Psi(\mathbf{p},t)
\end{eqnarray*}
\begin{equation}
\;\;\;\;\;\;\;\;\;\;\;\;\;\;\;\;\;\;\;\;\;\;\;\;\;\;\;-\int\frac{\mathrm{d}\mathbf{u}}{(2\pi)^3}V(\mathbf{p}-\mathbf{u})\Psi(\mathbf{u},t)=0.
\end{equation}
$\mathbf{p}$ is the canonical momentum and $V(\mathbf{p}-\mathbf{u})$ is the kernel of the Coulomb potential. The main idea of the model is to replace this kernel by a sum of $N$ symmetric separable potentials supporting $N$ bound states of atomic hydrogen:
\begin{equation}
V(\mathbf{p}-\mathbf{u})=-\frac{4\pi}{|\mathbf{p}-\mathbf{u}|^2}\approx V(\mathbf{p},\mathbf{u})=-\sum_{n=1}^Nv_n(\mathbf{p})v_n^*(\mathbf{u}).
\end{equation}
In order to solve Eq. (4) with the kernel of the Coulomb potential given by Eq. (5),  we first perform a contact transformation of the wave function $\Psi(\mathbf{p},t)$ to eliminate the $A^2(t)$ term from Eq. (4),
\begin{equation}
\Psi(\mathbf{p},t)=e^{-i\zeta(t)}\Phi(\mathbf{p},t),
\end{equation}
where, 
\begin{equation}
\zeta(t) = \frac{1}{2c^2}\int_0^tA(\xi)^2\mathrm{d}\xi.
\end{equation}
We then define the following function:
\begin{equation}
F_n(t)=\int\frac{\mathrm{d}\mathbf{u}}{(2\pi)^3}v^*_n(\mathbf{u})\Phi(\mathbf{u},t).
\end{equation}
Under these conditions, the solution of Eq. (4) with the kernel of the Coulomb potential replaced by the expression (5) may be formally written as follows:
\begin{eqnarray*}
\Phi(\mathbf{p},t)=e^{-\mathrm{i}\frac{p^2}{2}t+\mathrm{i}b(t)p_z}\left[\Phi(\mathbf{p},0)+\mathrm{i}\sum_{n=1}^Nv_n(\mathbf{p})\right.
\end{eqnarray*}
\begin{equation}
\left.\times\int_0^t\mathrm{d}\xi F_n(\xi)e^{\mathrm{i}\frac{p^2}{2}\xi-\mathrm{i}b(\xi)p_z}\right].
\end{equation}
Upon substituting expression (9) of $\Phi(\mathbf{p},t)$ into Eq. (8), we obtain a system of $N$ coupled linear 1-dimensional Volterra integral equations of the second kind which may be written in matrix form as:
\begin{equation}
\mathbf{F}(t)=\mathbf{F}_0(t)+\int_0^t\mathbf{K}(t,\xi)\mathbf{F}(\xi)\mathrm{d}\xi.
\end{equation}
$\mathbf{F}(t)$ is a vector of dimension $N$, the components of which are the $F_n(t)$ functions. $\mathbf{F}_0(t)$ is a vector of the same dimension which results from the contribution of the initial wave function $\Phi(\mathbf{p},0)$ present in Eq. (9). $\mathbf{K}(t,\xi)$ which is the kernel of the Volterra equation, is a $N\times N$ matrix (see \cite{Tetchou2011} for the details of the calculations). Consequently, the 4-dimensional TDSE is reduced to a system of $N$ coupled 1-dimensional Volterra integral equations depending only on time  which is a major advantage when the integration of the TDSE over the spatial coordinates requires a large grid or basis set.

\subsection{Water molecule}

To generalize the previous model to the case of the water molecule, we first generate the spatial part $\Phi_0(\mathbf{r})$ of the HOMO 1{\it b}$_1$  in the configuration space. It is obtained by geometry optimization with the GAMESS(US) program \cite{Schmidt1993} in the Hartree-Fock approximation. Instead of Hartree-Fock one could use DFT, it would not change much, as long as one uses experimental orbital energies instead of ones generated by GAMESS(US) as they are strongly influenced by the method while the wave functions are not.
The general expression of a molecular orbital $\alpha$ denoted by  $\Phi_{\alpha}(\mathbf{r})$ is:
\begin{equation}
\Phi_{\alpha}(\mathbf{r})=\sum_{j=1}^3\;\sum_{\gamma_j}C_{\gamma_j,\alpha}\;\mathcal{G}_{\gamma_j}(\mathbf{r}-\mathbf{R}_j),
\end{equation} 
where index $j$ designates each nucleus in the molecule. The se\-cond sum is over the atomic orbitals taken into account around each nucleus and $\mathcal{G}_{\gamma_j}$ is a so-called contracted gaussian from a 6-31G basis set \cite{Davidson1986} in the present case. $\mathbf{r}-\mathbf{R}_j$ is the electronic coordinate relative to the nucleus $j$. The coefficients $C_{\gamma_j,\alpha}$ are the ones generated by the GAMESS(US) program. The HOMO molecular orbital $\Phi_0(\mathbf{r})$ is then expressed in momentum space by Fourier transform. The corresponding separable potential 
\begin{equation}
V(\mathbf{p},\mathbf{u})=-v(\mathbf{p})v^*(\mathbf{u})
\end{equation}
is calculated by imposing that the molecular orbital $\Phi_0(\mathbf{r})$ is a solution of the following stationary Schr\"odinger equation:
\begin{equation}
\left[\varepsilon_0-\frac{p^2}{2}\right]\Phi_0(\mathbf{p})-\int\frac{\mathrm{d}\mathbf{u}}{(2\pi)^3}V(\mathbf{p},\mathbf{u})\Phi_0(\mathbf{u})=0,
\end{equation}
where $\varepsilon_0=-0.463$ a.u. is the experimental value of the energy of the HOMO 1{\it b}$_1$ \cite{Banna1975}. If we define the coefficient 
\begin{equation}
a=\int\frac{\mathrm{d}\mathbf{u}}{(2\pi)^3}v^*(\mathbf{u})\Phi_0(\mathbf{u}),
\end{equation}
we obtain immediately:
\begin{equation}
v(\mathbf{p})=-\frac{1}{a}\left[\varepsilon_0-\frac{p^2}{2}\right]\Phi_0(\mathbf{p}).
\end{equation}
The coefficient $a$ is easily obtained by multiplying Eq. (13) by $\Phi_0^*(\mathbf{p})$ and integrating over $\mathbf{p}$.
Consequently, the TDSE that describes, within the SAE approximation, the ejection of an electron from the HOMO 1{\it b}$_1$ of the water molecule exposed to a laser pulse is:
\begin{equation}
\left[\mathrm{i}\frac{\partial}{\partial t}-\frac{p^2}{2}+\frac{A(t)}{c}(\mathbf{e}\cdot\mathbf{p})\right]\Phi(\mathbf{p},t)-\int\frac{\mathrm{d}\mathbf{u}}{(2\pi)^3}V(\mathbf{p},\mathbf{u})\Phi(\mathbf{u},t)=0,
\end{equation}
with the initial condition:
\begin{equation}
\Phi(\mathbf{p},0)=\Phi_0(\mathbf{p}).
\end{equation}
As in the case of atomic hydrogen, the solution of this equation is obtained analytically:
\begin{equation}
\Phi(\mathbf{p},t)=e^{-\mathrm{i}p^2/2t+\mathrm{i}b(t)(\mathbf{e}\cdot\mathbf{p})}\left[\Phi_0(\mathbf{p})+\mathrm{i}v(\mathbf{p})\int_0^tF(\xi)e^{\mathrm{i}p^2/2\xi-\mathrm{i}b(\xi)(\mathbf{e}\cdot\mathbf{p})}\mathrm{d}\xi\right],
\end{equation}
and the function $F(t)$ is defined by:
\begin{equation}
F(t)=\int\frac{\mathrm{d}\mathbf{u}}{(2\pi)^3}v^*(\mathbf{u})\Phi(\mathbf{u},t).
\end{equation}
This function is the solution of the following 1-dimensional Volterra integral equation:
\begin{equation}
\label{eq_volt}
F(t)=F_0(t) + \int_0^tK(t,\xi)F(\xi)\mathrm{d}\xi.
\end{equation}
As we stressed before, this function depends only on time. All the spatial dependence is "hidden" in the kernel $K(t,\xi)$. In other words, the spatial dependence is treated analytically which allows one to avoid all the problems related to the size of the grid or of the basis used to integrate on the spatial coordinate of the electron. $F_0(t)$ and $K(t,\xi)$ can be expressed in terms of the following functions (see the Appendix for more details):
\begin{eqnarray}
\label{eq_STQ1}
S(x,y)&=&\int\frac{\mathrm{d}\mathbf{p}}{(2\pi)^3}\Phi^*_0(\mathbf{p})\Phi_0(\mathbf{p})e^{-\mathrm{i}p^2x+\mathrm{i}y(\mathbf{e}\cdot\mathbf{p})},\\
T(x,y)&=&\int\frac{\mathrm{d}\mathbf{p}}{(2\pi)^3}p^2\Phi^*_0(\mathbf{p})\Phi_0(\mathbf{p})e^{-\mathrm{i}p^2x+\mathrm{i}y(\mathbf{e}\cdot\mathbf{p})},\\
Q(x,y)&=&\int\frac{\mathrm{d}\mathbf{p}}{(2\pi)^3}p^4\Phi^*_0(\mathbf{p})\Phi_0(\mathbf{p})e^{-\mathrm{i}p^2x+\mathrm{i}y(\mathbf{e}\cdot\mathbf{p})}.
\end{eqnarray}
In the case of the HOMO 1{\it b}$_1$, we have:
\begin{equation}
F_0(t)=-\frac1a\left.\left[\varepsilon_0S(x,y)-\frac12T(x,y)\right]\right|_{x=t/2,y=b(t)},
\end{equation}
and
\begin{equation}
K(t,\xi)=\frac{\mathrm{i}}{a^2}\left[\varepsilon_0^2S(x,y)-\varepsilon_0T(x,y)+\frac14Q(x,y)\right]\left.\right|_{x=(t-\xi)/2,y=b(t)-b(\xi)}.
\end{equation}

The main shortcomings of the present method are the fact that the separable potential, which is unique if only one molecular orbital of a given symmetry is treated, does not necessarily have the correct asymptotic behaviour as well as the gauge invariance problem. The pro\-blem of the gauge invariance in the context of nonlocal potentials has been discussed in great detail in various references \cite{Starace1982,Rensink2016,Galstyan2017}. In particular, it is shown in \cite{Galstyan2017} that it is possible to formulate our se\-parable potential model in various ways that can be grouped into two families such that within a given family, the length and velocity gauge formulations give the same value for the obser\-vables. This shows clearly the non-existence of a privileged gauge but does not solve the problem of the "global" gauge invariance since formulations belonging to the first and the second family give different results for the observables. In the case of atomic hydrogen exposed to a relatively high frequency (of the order or higher than the ionisation potential) field, the present velocity gauge formulation of our model with several bound states taken into account gives results that are in good agreement with the TDSE results with the full Coulomb potential.\\

\subsection{Numerical implementation}
\label{sec_num}
In SPAM, Eq. (\ref{eq_volt}) is solved using the block-by-block method \cite{Linz1985}.  From the computational point of view, the most expensive step is the calculation of the kernels $K(t,\xi)$, and it strongly depends on the number of gaussians in the basis set. This number depends on the system under consideration and on the basis set. Calculation for atomic hydrogen with a huge basis set can be as expensive as a calculation for a water molecule, but with a small basis set.

However, the algorithm can be effectively parallelised. We use NVIDIA CUDA for General Purpose Graphics Processor Units (GPGPUs) combined with Message Passing Interface (MPI), so we use several nodes with GPUs. The overall time to calculate the ionisation yield at a high frequency for a water molecule using the 6-31G or 6-311++G** basis set \cite{Davidson1986} is around 30 minutes and 6 hours respectively on one GPU of CUDA compute capability 2.0. We use two GPUs with MPI to control the error during the calculation, or we use up to 64 GPUs simultaneously to compute an ionisation yield curve.

Having performed some calculations with small and large basis sets for the same parameters, we cannot find any significant difference between them except the computation time. Thus all the calculations that we present are performed with a small basis set 6-31G. \\

\section{Results and Discussions}

This section is divided in two subsections. In the first one, we present some tests of the adequacy of our model in the case of atomic hydrogen exposed to a XUV pulse. Results for the water molecule are presented and discussed in the second subsection.

\subsection{Verification of the model}

\begin{figure}[!h]
\centering
 \includegraphics[width=0.7\textwidth]{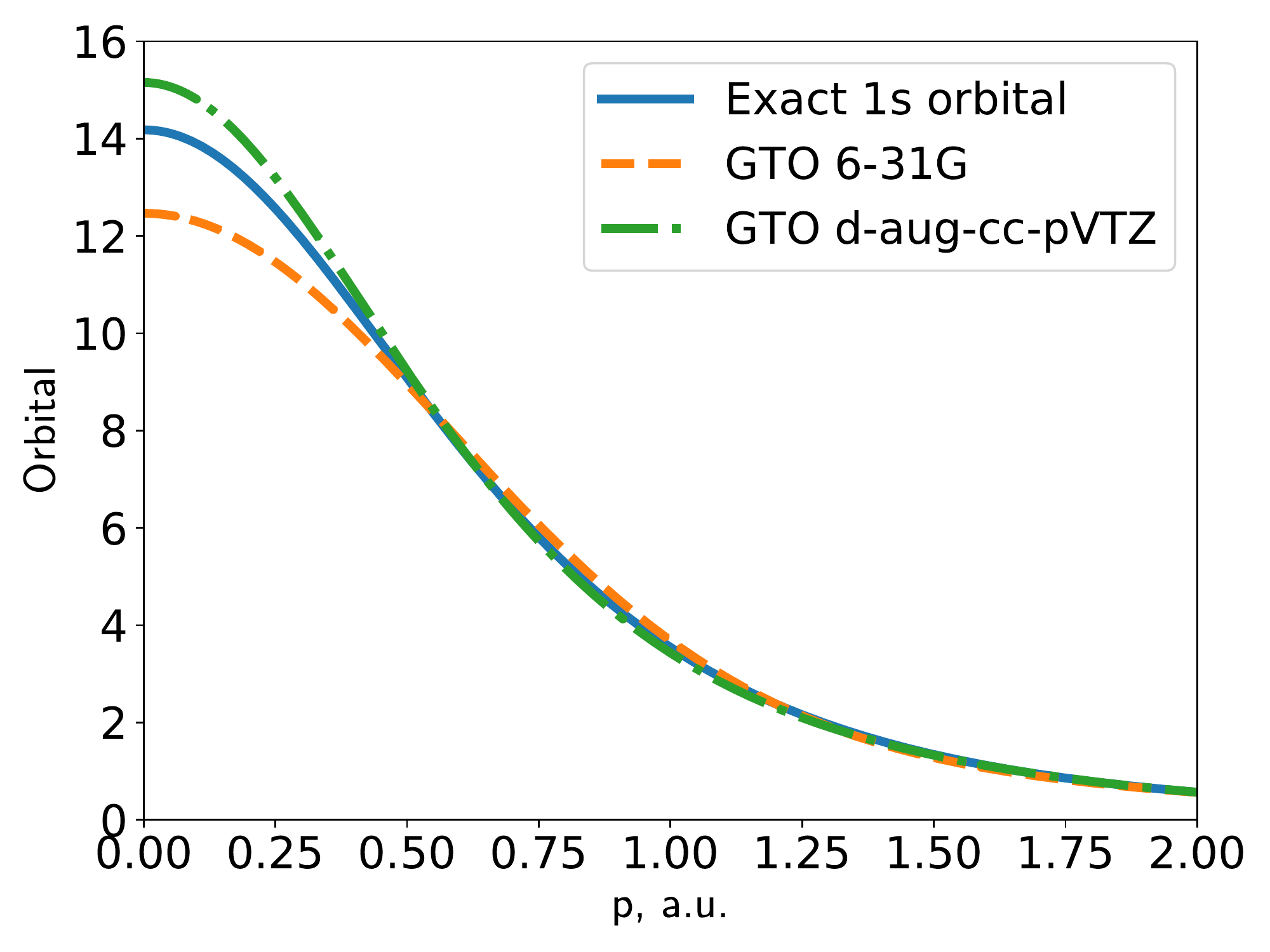}
 \caption{(Colour online) Comparison of the exact atomic hydrogen ground state wave function (radial part) in momentum space with the corresponding wave function (radial part) generated by the GAMESS(US) program in two different GTO basis sets.}
 \label{fig:Hydrogen1}
 \end{figure}

In the case of atomic hydrogen, the atomic state wave functions are known analytically. It is therefore interesting to analyze to what extent the same wave function can be accurately reproduced in a basis of gaussian type orbitals (GTO) with the GAMESS(US) program. In Fig. \ref{fig:Hydrogen1}, we compare the exact 1s-state wave function
 \begin{figure}[!h]
 \centering
  \subfloat[][SPAM (circles) and full TDSE (squares) results, compared with corresponding LOPT theories]{\includegraphics[width=0.65\textwidth]{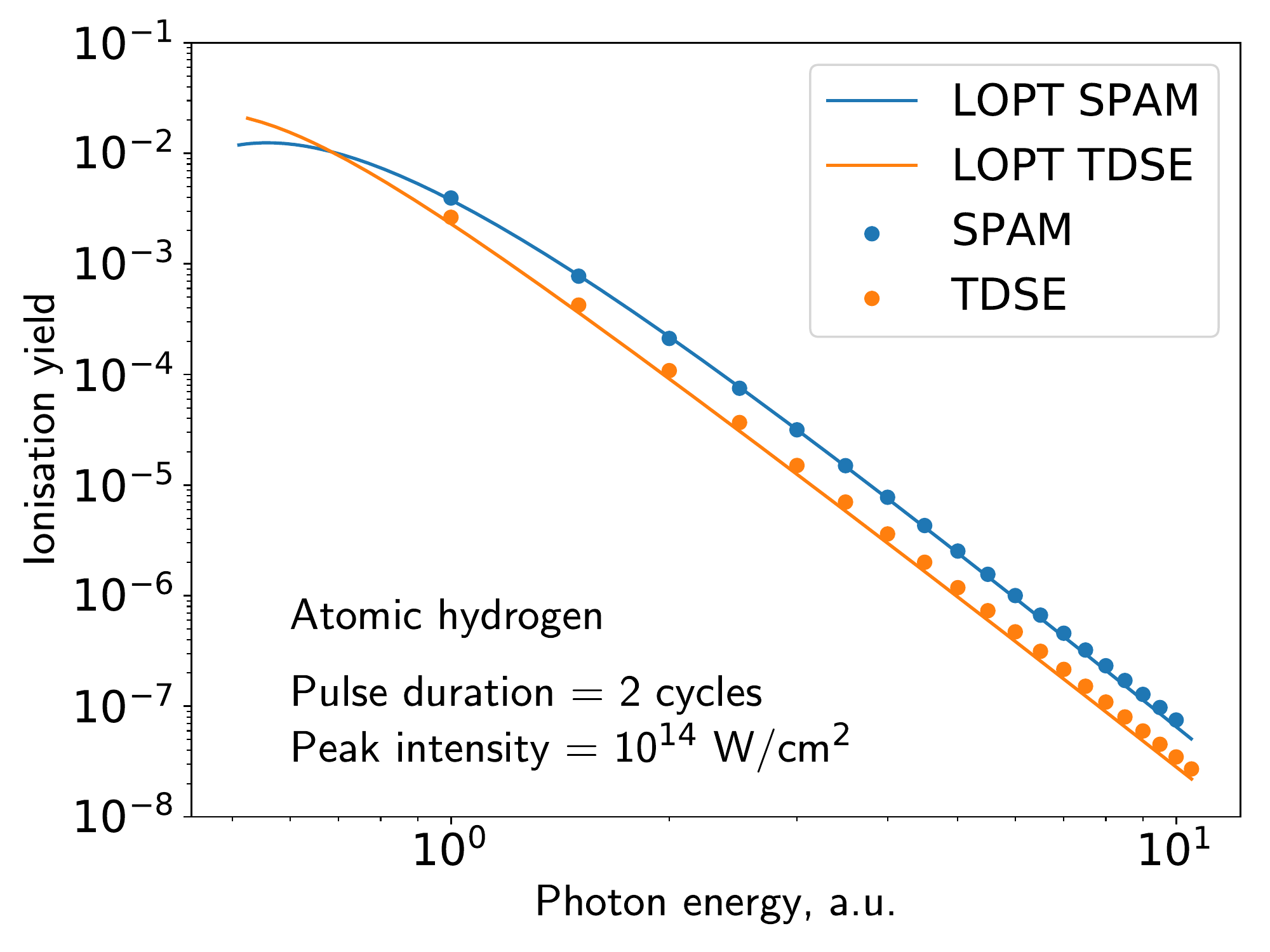}\label{fig:hydrogen_tdse_LOPT}}\\
 \subfloat[][SPAM (full line) and full TDSE (squares) results; SPAM result is scaled with a factor obtained from SPAM LOPT]{\includegraphics[width=0.65\textwidth]{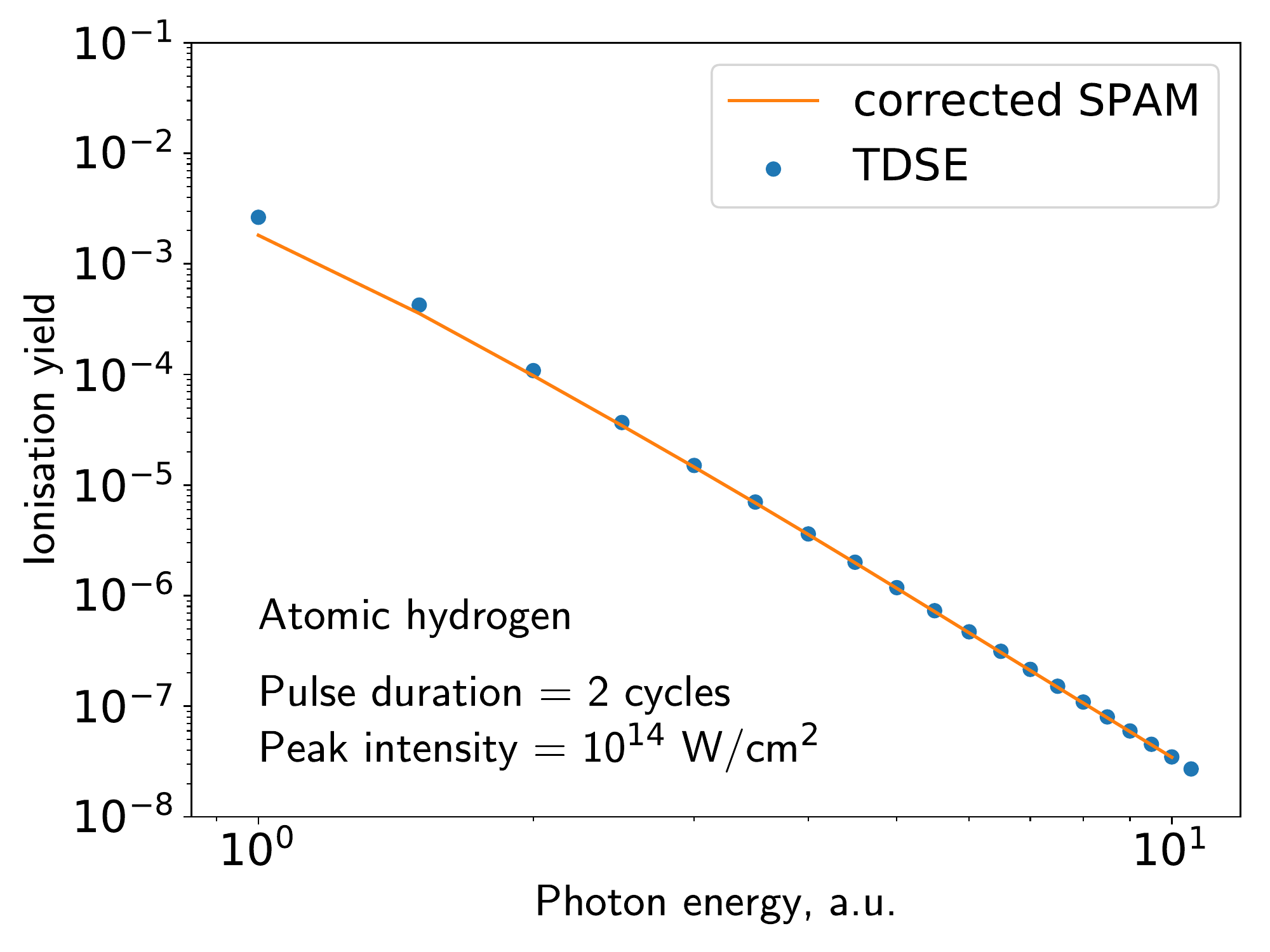}\label{fig:hydrogen_tdse_corrected}}
 \caption{(Colour online) Ionisation yield as a function of the photon energy for atomic hydrogen exposed to a two-cycle pulse with a carrier envelope phase equal to zero and 10$^{14}$ W/cm$^2$ peak intensity.}
\end{figure}
(radial part) in momentum space with the corresponding wave function (radial part) generated by the GAMESS(US) program in two different bases of GTO's, a small one (6-31G) and a bigger one (6-311++G**). A correlation consistent basis set d-aug-cc-pVTZ has also been used but the resulting radial part of the wave function which coincides with the one obtained with the 6-311++G** basis set, is not shown in the figure. These results show clearly that the GTO basis sets are not accurate in the region of  small momenta. This is expected since small momenta correspond, in the configuration space, to large distances which are not well described by gaussians that are very localized functions. Although the small distances can be described relatively well using a small basis set (see the Figure \ref{fig:Hydrogen1}), we have to account for the long range behaviour as well. In this case, a bigger basis set works better. 

In Fig. \ref{fig:hydrogen_tdse_LOPT}, we show the ionisation yield as a function of the frequency in the case where atomic hydrogen is exposed to a two-cycle pulse of zero carrier envelope phase and for 10$^{14}$ W/cm$^2$ peak intensity. The calculations have been performed in two ways: (i) by solving the TDSE with the Coulomb potential fully taken into account and (ii) by using our model in which we only take into account the ground state of atomic hydrogen. In the latter case, describing the atomic ground state with the exact wave function or in a basis of GTO's does not affect the results despite the fact that GTO's are unable to reproduce accurately the behaviour of the wave function for small momenta. We clearly see in Fig. \ref{fig:hydrogen_tdse_LOPT} that the results obtained by solving TDSE and by using our model are very close to each other, although with a constant shift. To unveil the nature of this shift we performed two different lowest order perturbation theory (LOPT) calculations. In the first LOPT calculation, we considered single photon ionisation from the ground state of hydrogen atom to a continuum p-wave, described by a Coulomb wave. This is the LOPT TDSE curve, that coincides perfectly with the result of TDSE calculation. The second LOPT calculation has a plane wave as the final state, and everything else is the same as in the first case. This is the LOPT SPAM curve, that coincides perfectly with the SPAM calculation. We can conclude that the constant shift that we observe is due to the wrong description of the continuum, which is close to a plane wave in the case of SPAM model, while it is a Coulomb wave in reality. However, what is hidden behind the log-log scale of the plot is the fact that the absolute difference between the SPAM and TDSE becomes smaller and smaller. This shows that for high frequencies we can approximate a Coulomb wave with a plane wave. It turns out that using plane wave instead of Coulomb wave is equivalent to introducing a constant factor. We use this factor to correct the predictions for atomic hydrogen and for water molecule as well. In fact, this factor is not "ad hoc" as it can be derived from the dipole matrix elements in the case of a Coulomb wave and a plane wave final state of the ionising system. The comparison of the corrected SPAM result with full TDSE result is given in Figure \ref{fig:hydrogen_tdse_corrected}. 

\subsection{Water molecule}

Let us now consider the water molecule. In this case, full TDSE calculations are not tractable and must rely on some approximation schemes. In the high frequency regime considered here and given the good agreement we obtained between TDSE and our model calculations for atomic hydrogen, it is reasonable to think that applying our approach to the water molecule considered as a single active electron system should give relatively good results. 

In this subsection, we study the probability of orientation-dependent single ionization of the HOMO 1{\it b}$_1$ by a laser pulse for various frequencies, peak intensities and carrier envelope phases. The reason why we chose water instead of a simpler multielectron molecule like H$_2$ is that the water molecule is a very convenient multielectron quantum system for our purposes as: (a) its HOMO orbital is essentially a 2p atomic orbital of the oxygen atom with very little influence of each hydrogen atom (the value of magnetic quantum number for this 2p orbital depends on the particular position of the molecule in the reference frame); (b) ionisation of the water HOMO orbital leaves the geometry of the molecule unchanged, thus allowing us to apply the fixed nuclei approximation as the vibrational excitation is low.

Since the full final momentum wavepacket is available in the SPAM model, any information about the system can be easily extracted. The absence of the intermediate states means however that regimes where these states are important, like low frequency ionisation, cannot be treated accurately. Being a SAE approach one neglects all the dynamic interactions between the particles. Finally the Born Oppenheimer approximation neglects all the processes related to motion of the nuclei. Nevertheless, the SPAM model allows one to make predictions for any complex system, where the aforementioned approximations are adequate, in the single photon regime. The model is very scalable, so the limits on the size of the system are given by the hardware resources.

\begin{figure}[!h]
\centering
\subfloat[][HOMO parallel to {\it z}-axis]{\includegraphics[width=0.45\textwidth]{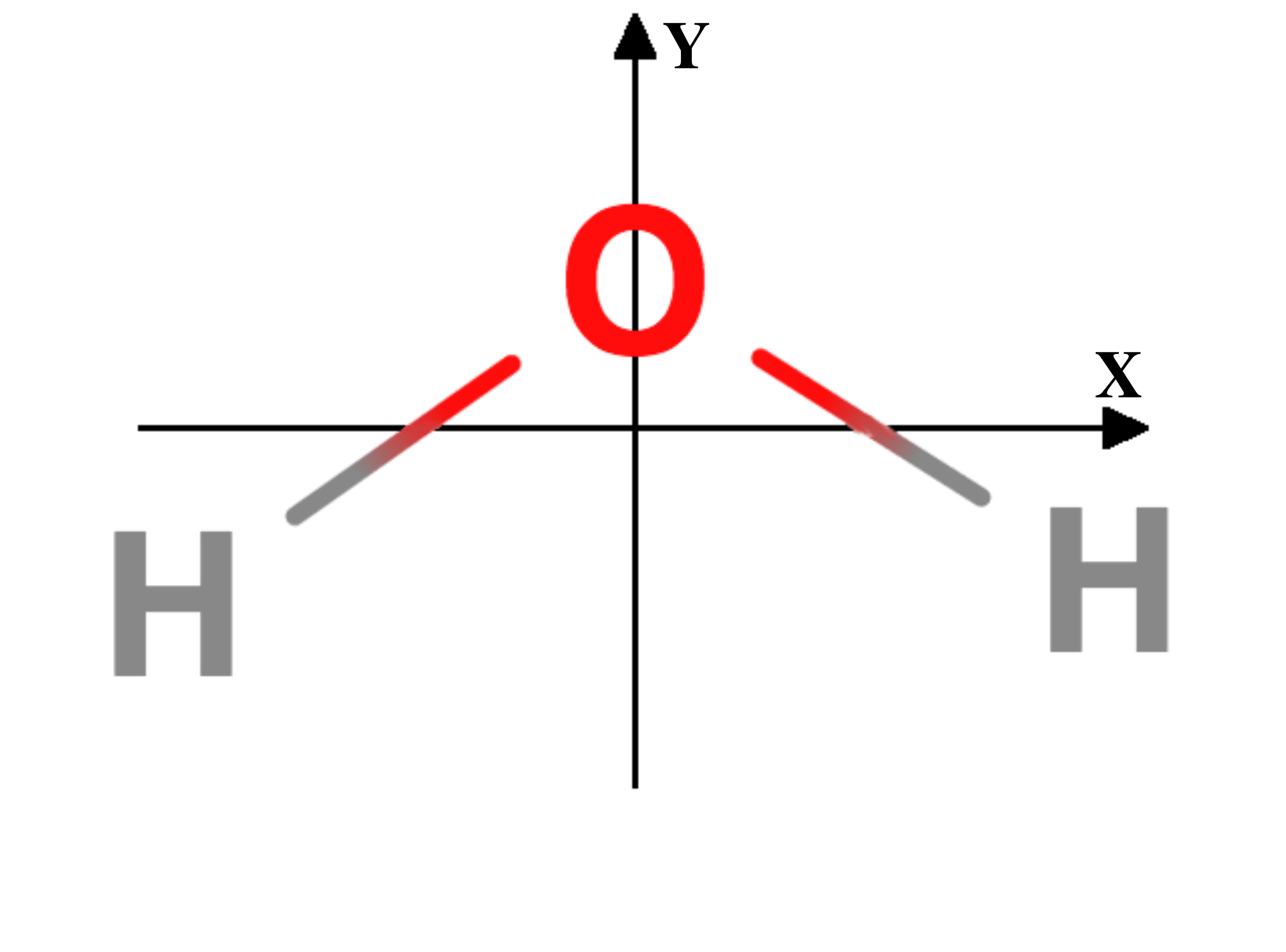}\label{fig:geometry_yx}}
\subfloat[][HOMO parallel to {\it x}-axis]{\includegraphics[width=0.45\textwidth]{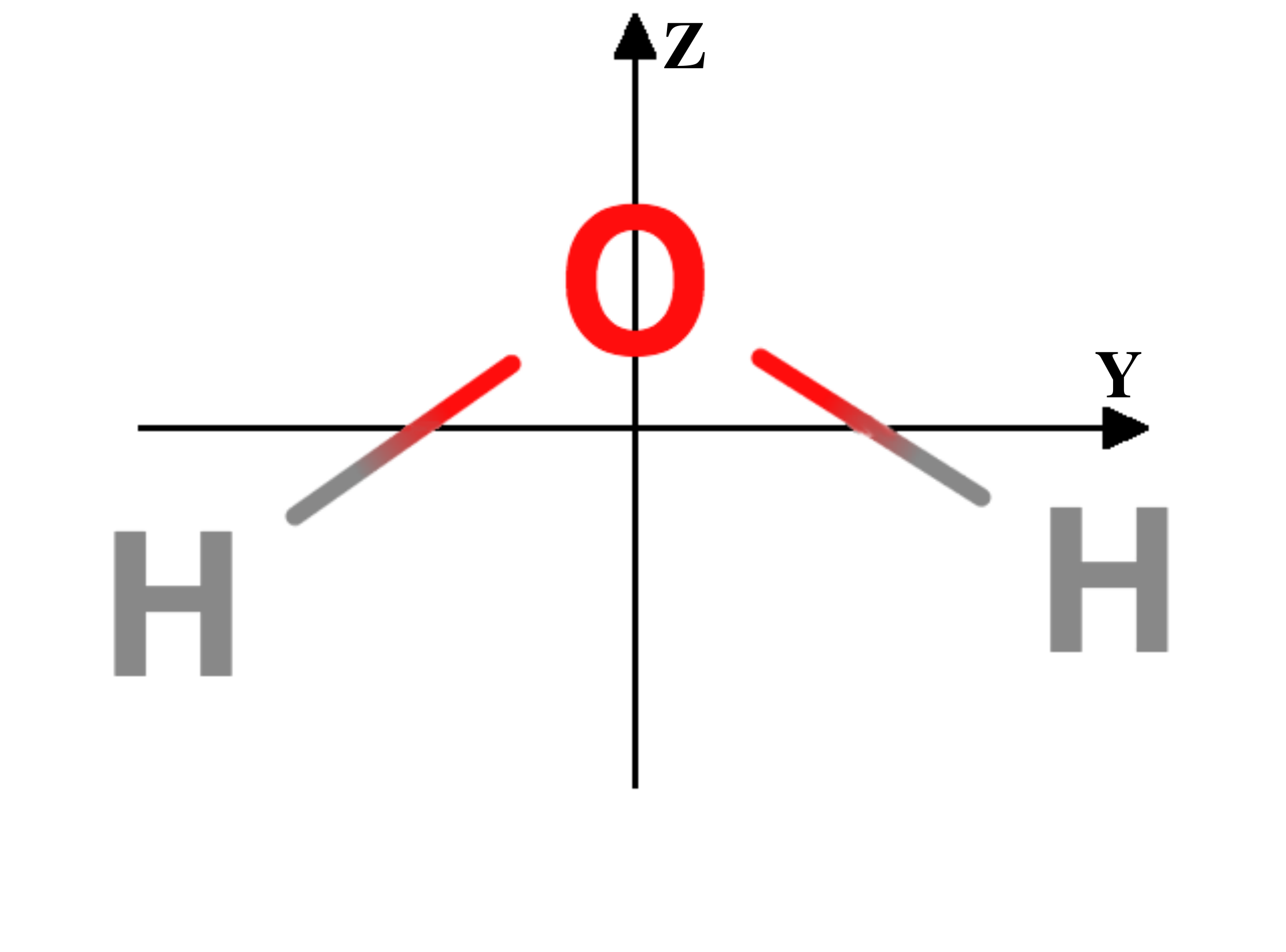}\label{fig:geometry_yz}}
\caption{(Colour online) Position of the water molecule in space. The origin of the reference frame is in the molecular plane. The HOMO orbital is perpendicular to the molecular plane.}
\label{fig:geometry}
\end{figure}

Fig. \ref{fig:geometry} shows two possible orientations of the water molecule with respect to the reference frame that are used later for the calculations. The origin of this reference frame is in the molecular plane. Let us recall that for all the cases treated here the electric field is polarized along the z-axis.
All the calculations have been performed in the 6-31G basis set. We are not aware of any significant discrepancies in our calculations that could be attributed to the incompleteness of this basis.\\

\begin{figure}[!h]
 \centering
 \subfloat[][Polarization vector parallel to the z-axis]{\includegraphics[width=0.55\textwidth]{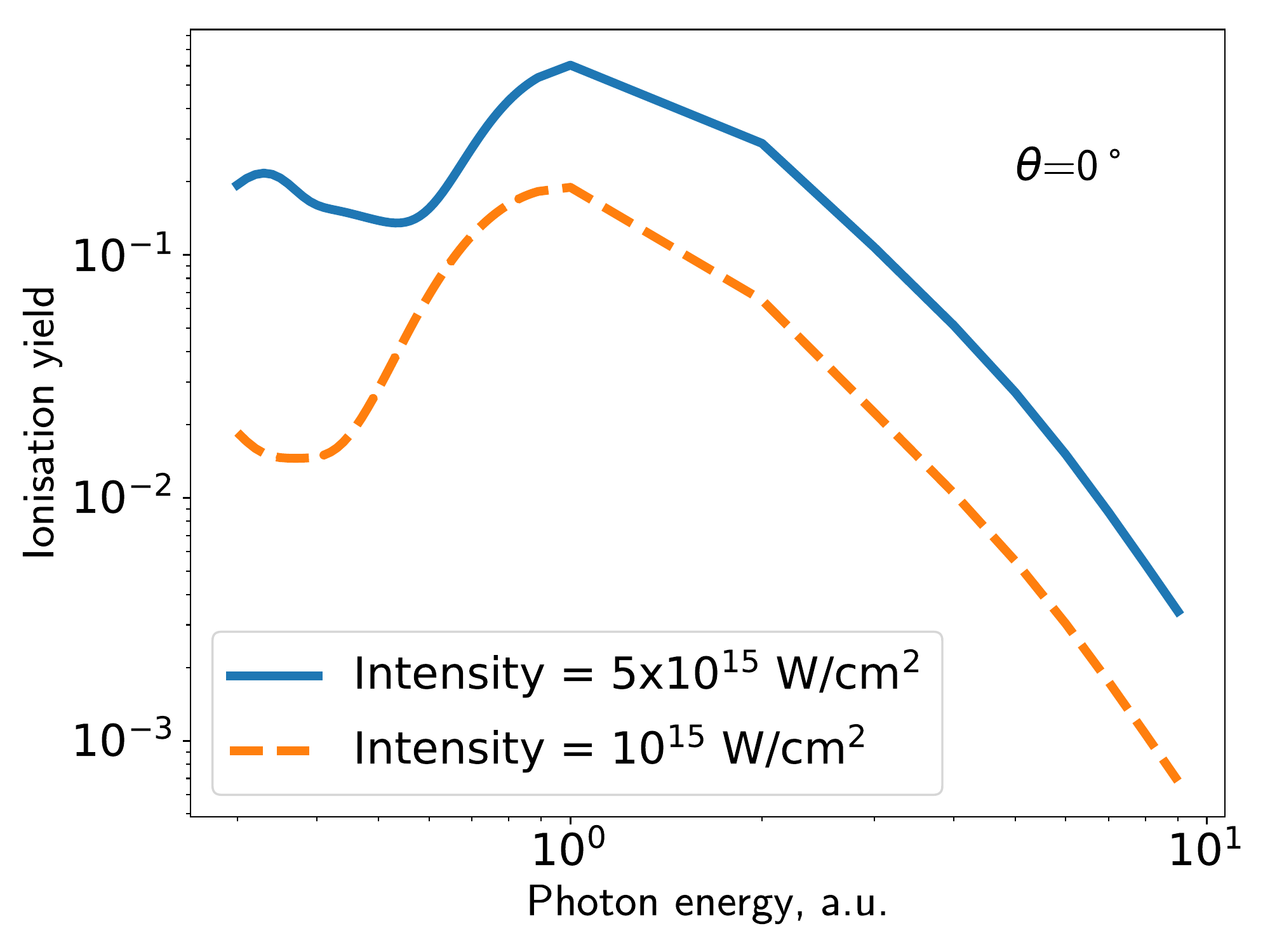}}
 \subfloat[][Polarization vector turned by $30^{o}$ in the xz-plane]{\includegraphics[width=0.55\textwidth]{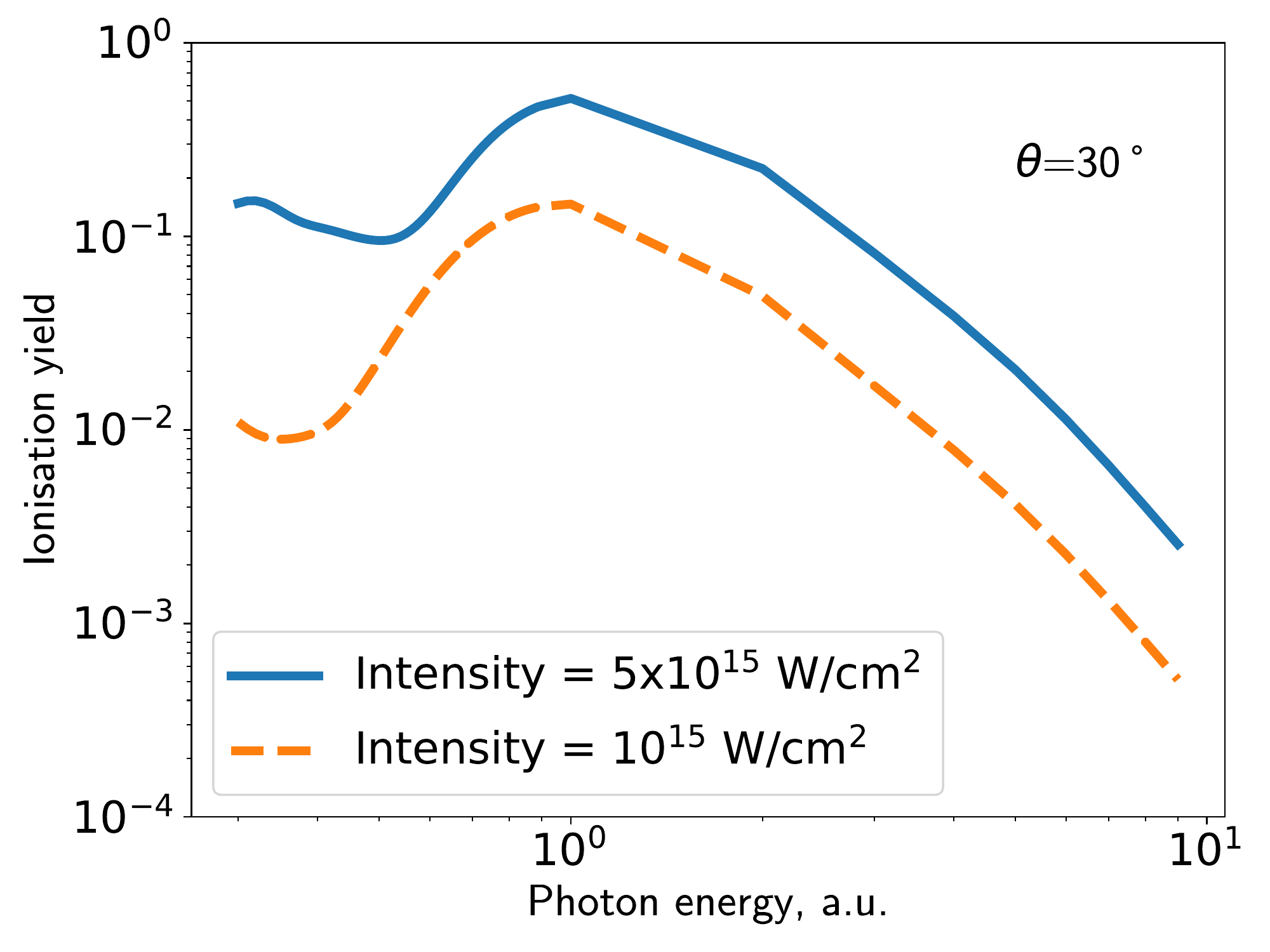}}\\
 \subfloat[][Polarization vector turned by $60^{o}$ in the xz-plane]{\includegraphics[width=0.55\textwidth]{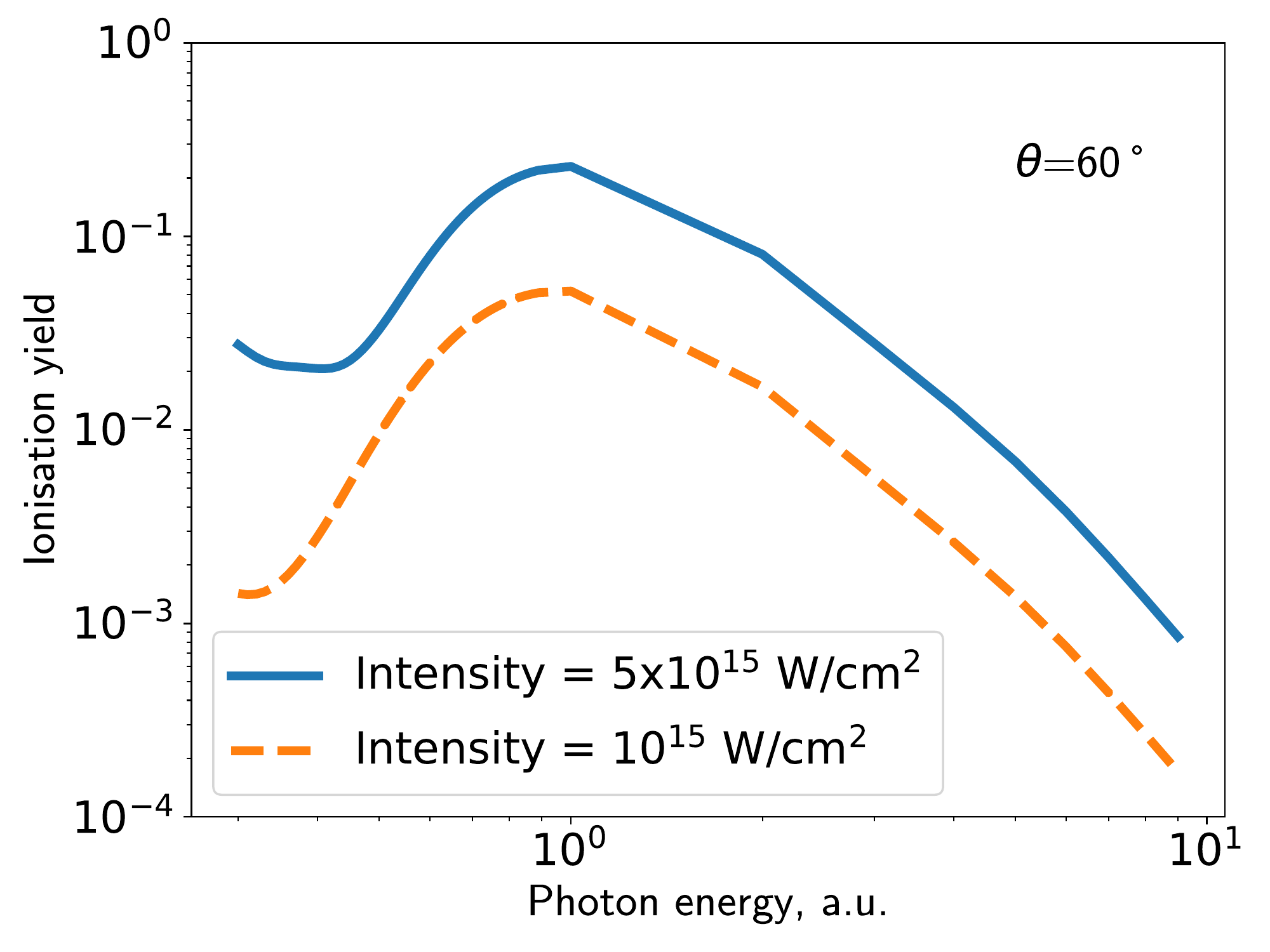}}
 \caption{(Colour online) Geometry like in Fig. \ref{fig:geometry_yx}. Dependence of the ionisation yield of water molecule on the photon energy for a pulse duration of 900 as. Two peak intensities are considered: $5\times10^{15}$ W/cm$^2$ (solid blue line) and $10^{15}$ W/cm$^2$ (dashed orange line). The result is not multiplied by the correction factor.}
 \label{fig:photon_energy}
\end{figure}

\begin{figure}[!h]
 \centering
 \subfloat[][Laser pulse frequency 0.3 a.u.]{\includegraphics[width=0.55\textwidth]{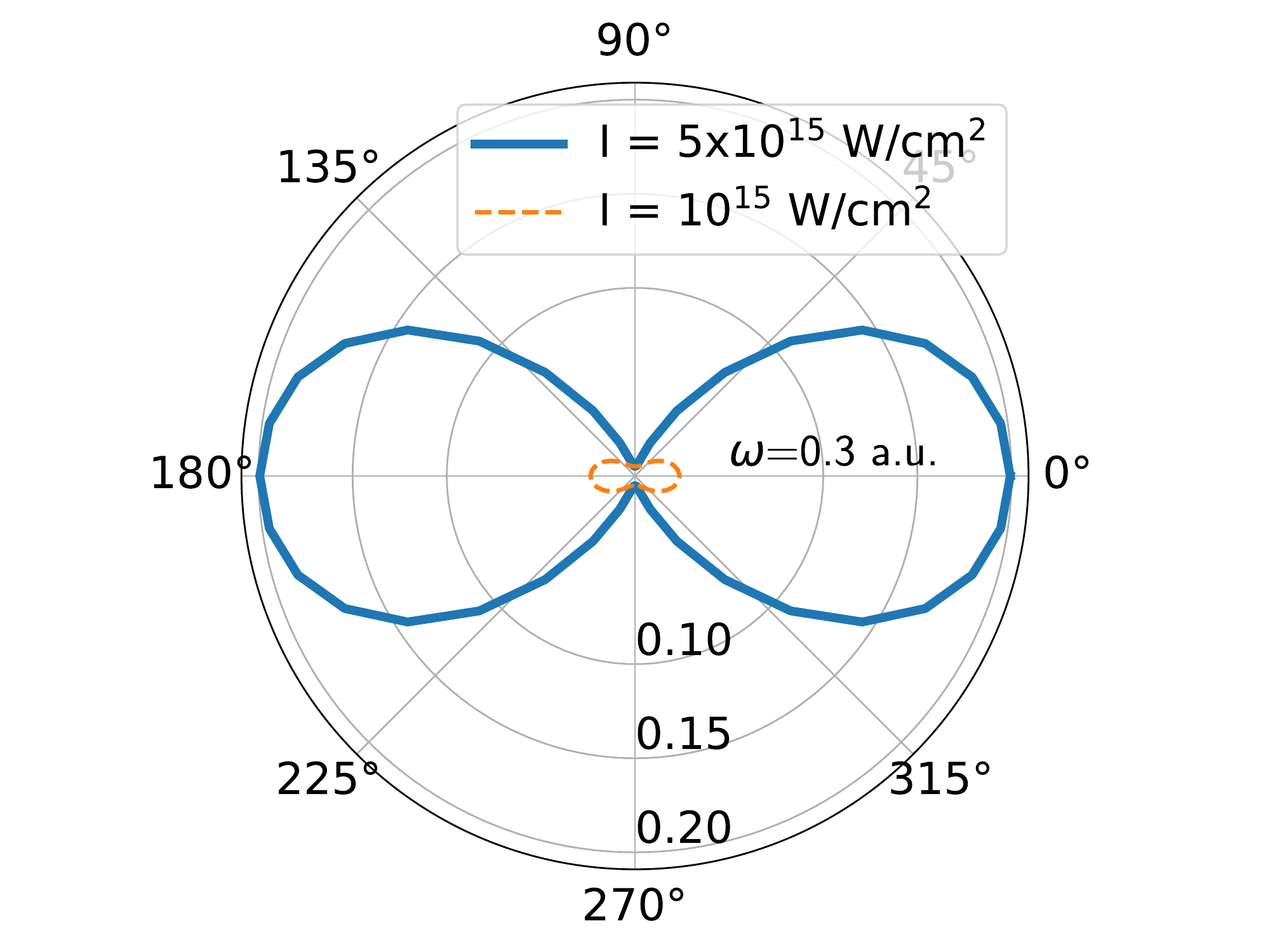}}
 \subfloat[][Laser pulse frequency 0.5 a.u.]{\includegraphics[width=0.55\textwidth]{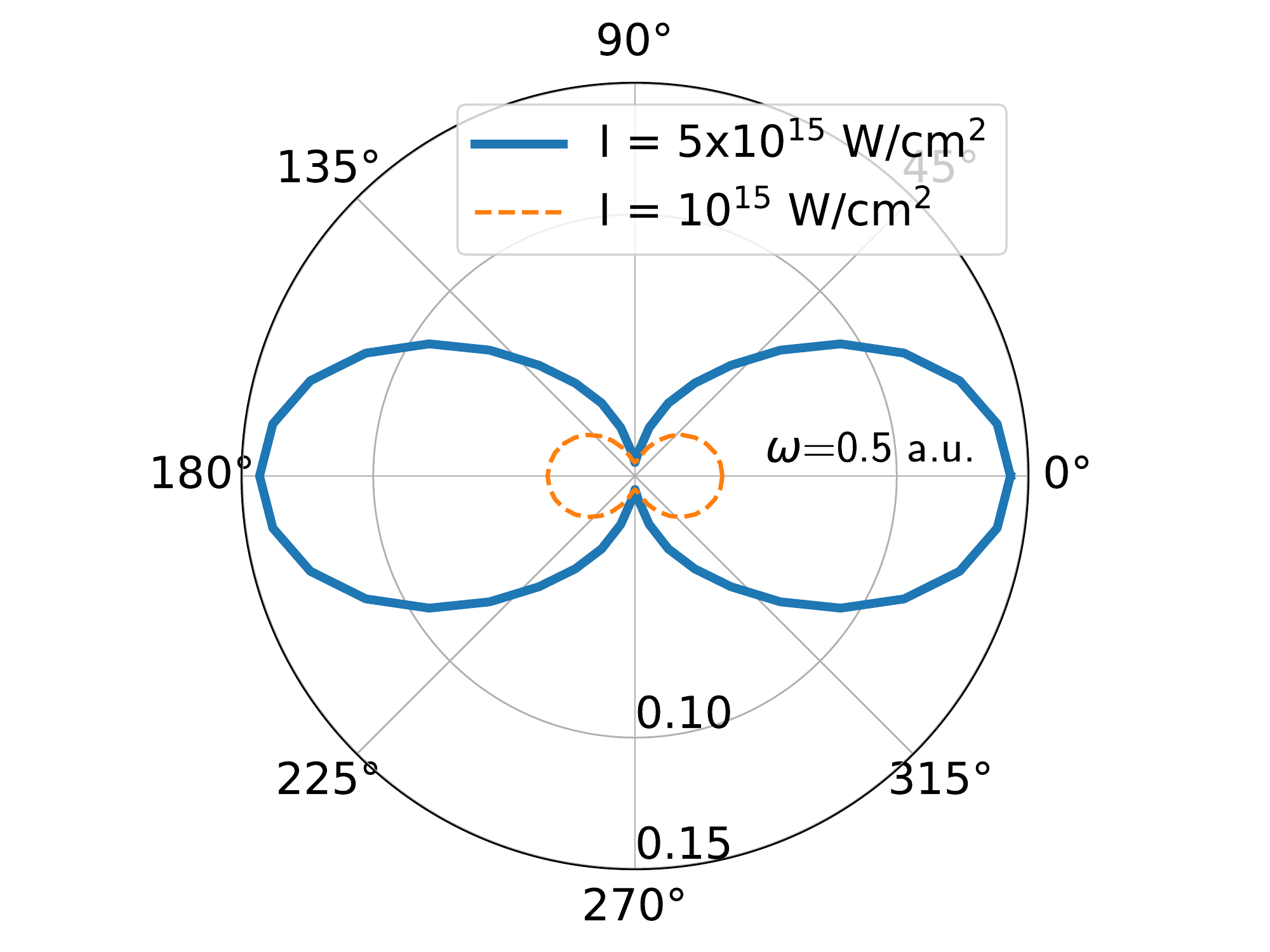}}\\
 \subfloat[][Laser pulse frequency 1 a.u.]{\includegraphics[width=0.55\textwidth]{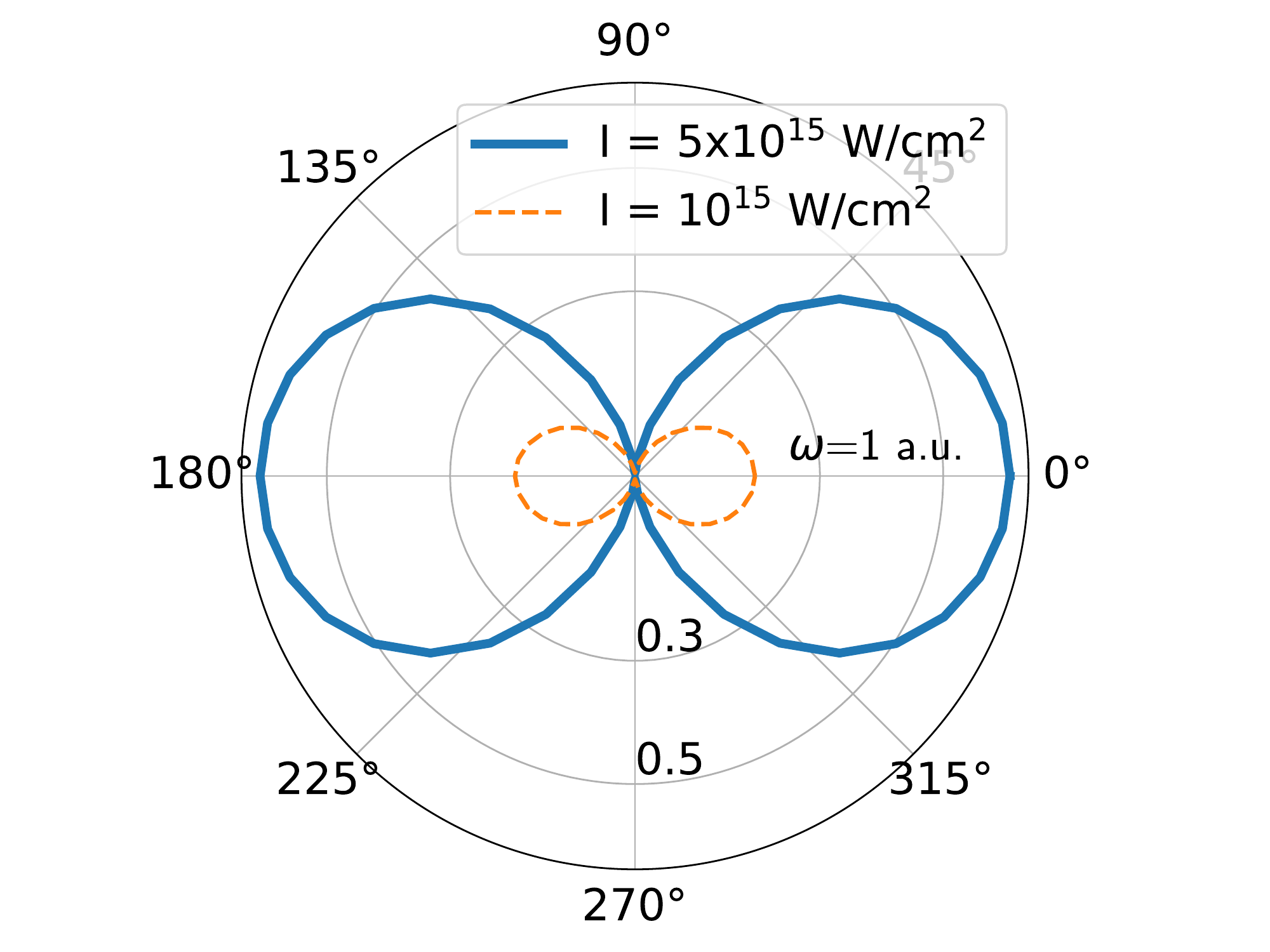}}
 \caption{(Colour online) Geometry like in Fig. \ref{fig:geometry_yx}. Dependence of the ionisation yield on the angle $\theta$ between the polarization direction in the xz-plane and the z-axis for a water molecule exposed to 900 as pulses of three different frequencies. Two peak intensities are considered:  $5\times10^{15}$ W/cm$^2$ (solid blue line) and $10^{15}$ W/cm$^2$ (dashed orange line). Some parts of these figures were obtained using the symmetry of the system. The result is not multiplied by the correction factor.}
 \label{fig:polarization}
\end{figure}

In Fig. \ref{fig:photon_energy}, we consider the photon energy dependence of the probability of ionisation of the water molecule placed as in Fig. \ref{fig:geometry_yx}, by a laser pulse with a fixed duration of 900 as, for three different angles between the polarization vector and the {\it z}-axis and for two peak intensities: $10^{15}$ W/cm$^2$ and $5\times10^{15}$ W/cm$^2$. It can be seen that for a fixed peak intensity, the ionisation yield is the highest when the field polarization axis is along the axis of $2p_z$ oxygen orbital which practically coincides with the H$_2$O HOMO orbital. The yield decreases rapidly when the polarization vector is rotated towards the {\it x}-axis. This conclusion supports the findings of Petretti {\it et al.}\cite{Petretti2013}. They showed  that for a 800 nm wave length, the shape of the water molecule ionisation yield as a function of the orientation of the polarization vector is more or less the same as the shape of the molecular orbital from which the ionisation occurs. To stress this point, we show in Fig. \ref{fig:polarization}, the ionisation yield
as a function of the angle $\theta$ between the polarization direction in the {\it xz}-plane and the {\it z}-axis. When the polarization axis is perpendicular to the {\it z}-axis (therefore in the {\it xy}-plane), we obtain an ionization yield which is equal to zero within the machine accuracy. This contrasts with Petretti's results who obtained a small but finite yield for a frequency 10 times lower. Note that the ionisation yield gets flatter around $\theta=\pi/2$ for the lowest frequency, and also the overall shape shows significant variation.  \\

\begin{figure}[!h]
 \centering
 \includegraphics[width=0.7\textwidth]{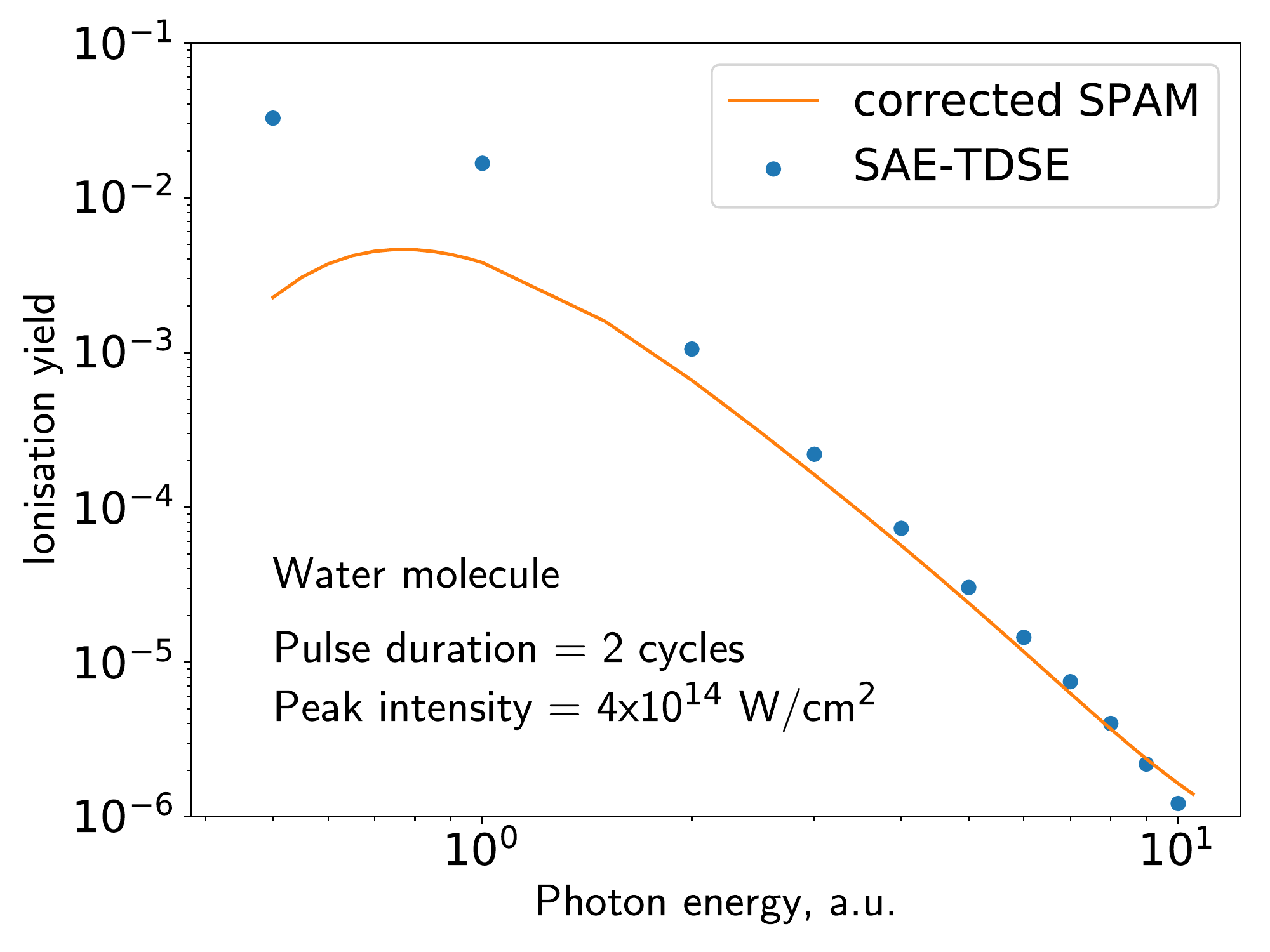}
 \caption{(Colour online) Geometry like in Fig. \ref{fig:geometry_yz}. Dependence of the ionisation yield of two water molecule models on the photon energy for a sine squared pulse of 2 cycle duration and $4\times10^{14}$ W/cm$^2$ peak intensity. The solid line has been obtained by using our SPAM method, the dots show the results of SAE-TDSE calculation \cite{Petretti2013}. }
 \label{fig:SAE_H2O_frequency}
\end{figure}

\begin{figure}[!h]
 \centering
  \includegraphics[width=0.7\textwidth]{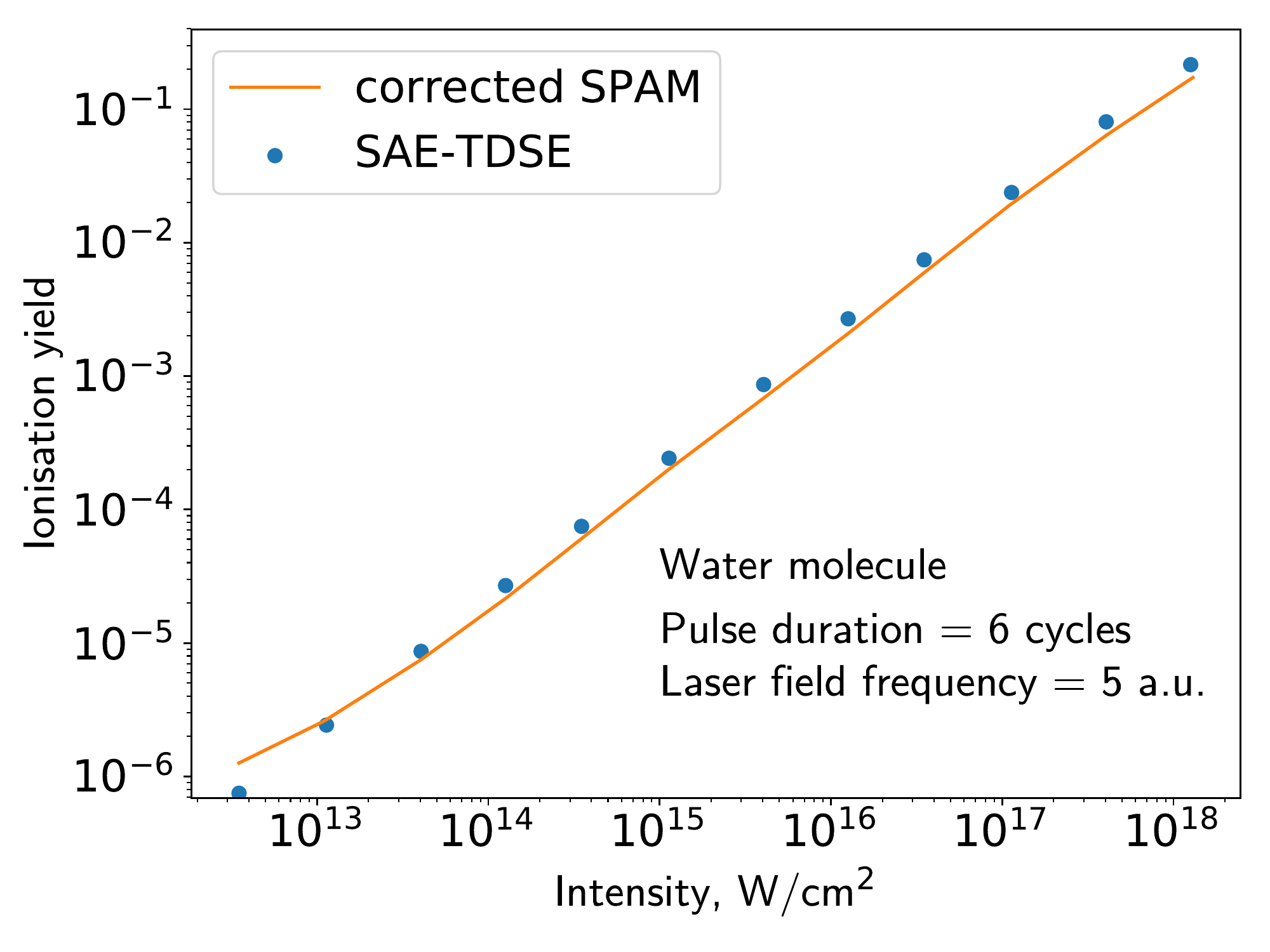}
 \caption{(Colour online) Geometry like in Fig. \ref{fig:geometry_yz}. Dependence of the ionisation yield of two water molecule models on the laser pulse peak intensity for a sine squared pulse of 6 cycle duration and $5$ a.u. photon energy. The solid line has been obtained by using our SPAM method, the dots show the results of SAE-TDSE calculation \cite{Petretti2013}. }
 \label{fig:SAE_H2O_intensity}
\end{figure}

In order to have some idea about how accurate is the prediction of the SPAM model for high frequencies, we ran the SAE-TDSE code from \cite{Petretti2013} in the 1-photon ionisation regime and compared the ionisation yield prediction of these two SAE models (see Figure \ref{fig:SAE_H2O_frequency}). A significant difference between these models is the fact that for SAE-TDSE the full coulomb potential has been taken into account in generating the orbital basis, thus we have to correct the SPAM result with the same factor as for atomic hydrogen. Both ionisation yields are normalised to 1. In fact, we have perfect agreement between these models for high frequencies, and poor agreement for the photon energies near the ionisation threshold. This can be explained by the fact that SAE-TDSE uses 6000 Kohn-Sham orbitals to propagate the wavefunction, while in SPAM we don't have any intermediate state at all. This discrepancy can be reduced as we include the lowest unoccupied orbital in the SPAM calculation, which will be done in future publications.

We can see in Figure \ref{fig:SAE_H2O_intensity} that the corrected SPAM coincides with SAE-TDSE in a wide intensity range as well. This agreement of the corrected SPAM model and SAE-TDSE approach in a wide intensity and frequency range indicates that the SPAM correction factor does not depend on intensity and on frequency.

\section{Conclusions and perspectives}

We have extended a new computationally inexpensive model, previously developed for atomic hydrogen to the treatment, within the SAE approximation, of the interaction of a complex quantum system with a high frequency ultrashort laser pulse. As a first application, we have applied this model to the single ionisation of the HOMO of the water molecule by an ultrashort XUV pulse. We studied the dependence of the single ionisation yield on the pulse frequency, the peak intensity, the orientation of the polarization vector and the carrier envelope phase. Our results clearly show that  the model  allows one to perform such calculations quickly for a complex system. Although it has been shown that this approach works rather well for atomic hydrogen in the low frequency regime we do not expect this model to be able to treat the ionisation of H$_2$O by a low frequency laser pulse because the intermediate states for a many electron quantum system are very important. However we will try to investigate the behaviour of the model water molecule in a laser pulse having HOMO and LUMO orbitals included.\\ \\

\begin{acknowledgement}

The authors thank Prof. Cl\'ement Lauzin for many useful discussions. A.G. is "aspirant au Fonds de la Recherche Scientifique (F.R.S.-FNRS)". Yu.P. thanks the Universit\'e catholique de Louvain (UCL) for financially supporting several stays at the Institute of Condensed Matter and Nanosciences of the UCL. F.M.F. and P.F.O'M. gratefully acknowledge the European network COST (Cooperation in Science and Technology) through the Action CM1204 "XUV/X-ray light and fast ions for ultrafast chemistry" (XLIC) for financing several short term scientific missions at UCL. P.D. and A.G. acknowledge COST XLIC and F.R.S-FNRS for financing two short term scientific missions (STSM) in Trieste, Italy, and participation in COST XLIC meetings. The present research benefited from computational resources made available on the Tier-1 supercomputer of the Federation Wallonie-Bruxelles funded by the Region Wallonne under the Grant No. 1117545 as well as on the supercomputer Lomonosov from Moscow State University and on the supercomputing facilities of the UCL and the Consortium des Equipements de Calcul Intensif (CECI) en Federation Wallonie-Bruxelles funded by the F.R.S.-FNRS under the convention 2.5020.11. Y.P. is grateful to the Russian Foundation for Basic Research (RFBR) for financial support under the grant No. 16-02-00049-a.\\ \\
\end{acknowledgement}





\providecommand{\latin}[1]{#1}
\makeatletter
\providecommand{\doi}
  {\begingroup\let\do\@makeother\dospecials
  \catcode`\{=1 \catcode`\}=2 \doi@aux}
\providecommand{\doi@aux}[1]{\endgroup\texttt{#1}}
\makeatother
\providecommand*\mcitethebibliography{\thebibliography}
\csname @ifundefined\endcsname{endmcitethebibliography}
  {\let\endmcitethebibliography\endthebibliography}{}

\end{document}